\documentclass[notitlepage,aps,11pt,amsmath,amsfonts,amssymb,nofootinbib]{revtex4-1}

\usepackage{enumerate}
\usepackage{graphicx}
\usepackage{amsmath}
\usepackage{amssymb}
\usepackage{enumerate}
\usepackage{graphicx}

\begin{document}



\title{The tangent bundle exponential map and locally autoparallel coordinates for general connections on the tangent bundle with application to Finsler geometry}

\author{Christian Pfeifer}
\email{christian.pfeifer@itp.uni-hannover.de}
\affiliation{Institute for theoretical physics,Leibniz Universit\"at Hannover, Appelstrasse 2, 30167 Hannover, Germany}

\begin{abstract}
We construct a tangent bundle exponential map and locally autoparallel coordinates for geometries based on a general connection on the tangent bundle of a manifold. As concrete application we use these new coordinates for Finslerian geometries and obtain Finslerian geodesic coordinates. They generalise normal coordinates known from metric geometry to Finsler geometric manifolds and it turns out that they are identical to the Douglas-Thomas normal coordinates introduced earlier. We expand the Finsler Lagrangian of a Finsler spacetime in these new coordinates and find that it is constant to quadratic order. The quadratic order term comes with the non-linear curvature of the manifold. From physics these coordinates may be interpreted as the realisation of an Einstein elevator in Finslerian spacetime geometries.
\end{abstract}

\maketitle


\section{Introduction}

Locally autoparallel coordinates respectively locally geodesic coordinates constructed with help of the exponential map are an important tool in metric and affine connection geometry for various proofs, calculations and derivations. Their most valuable properties are that the connection coefficients vanish at the origin of the coordinate system and that autoparallels of the affine connection respectively geodesics become straight lines. In gravitational physics the existence of coordinates which are optimally adapted to the Lorentzian metric geometry of spacetime, Riemann normal coordinates, is interpreted as the realization of the Einstein equivalence principle. Locally, in a small neighborhood around a physical observer on spacetime, the observer can neglect gravitational effects up to a certain distance to its position. This statement manifests mathematically in the fact that in Riemann normal coordinates around an observer's spacetime position the metric components depend only in second order on these coordinates \cite{Stephani}.

The most straightforward generalisation of this framework of affine connection geometry, for physics and mathematics, is obtained replacing the affine connection, which defines the geometry of the manifold, by a general connection on the tangent bundle or considering Finslerian geometry instead of metric geometry. Recently in particular Finsler geometry has drawn some attention in physics as possible extended non-metric geometry of spacetime which describes effects from classical gravity and quantum gravity phenomenology~\cite{Pfeifer:2011xi,Minguzzi2,Gibbons:2007iu,Girelli:2006fw,Amelino-Camelia:2014rga}. 

For Finslerian geometries it is known that there exist no smooth normal coordinates. Theorems on the non-existence of smooth normal coordinates are for example presented by Busemann \cite{Busemann}, Rund \cite{Rund} or Akbahr Zadeh \cite{Zadeh} (see \cite{BCS} for further references). The statement of the non existence theorems is that locally autoparallel coordinates respectively locally geodesic coordinates do not exist as coordinate system on the manifold in consideration. However it has been shown by Douglas and Thomas \cite{Douglas1927,Thomas1936} and Whitehead \cite{Whitehead1935} that on Finslerian geometries there exist direction dependent normal coordinates that are smooth everywhere except along the zero direction.

In this article we will demonstrate that for geometries based on general connections on the tangent bundle there exist coordinates on the tangent bundle in which the connection coefficients of the general connection vanish at the center point of the coordinate system on the manifold. For homogeneous and symmetric connections the general coordinates we construct become the Douglas-Thomas normal coordinates and the autoparallels of the non-linear connection become straight lines. We apply the general construction to Finslerian geometries 
and expand the geometry defining Finsler Lagrangian, a slight generalisation of the Finsler function, in a power series in the new position coordinates and obtain that it depends on them only from second order on, similarly as the components of the metric in normal coordinates in metric geometry. Therefore this is the starting point for physics, to investigate if the existence of the generalised locally autoparallel coordinates can be interpreted as a realisation of a Finslerian version of the Einstein equivalence principle in Finslerian spacetime geometries.

The key ingredient, and difference to the earlier approaches, in the construction of the general normal coordinates is that we first construct 
coordinates directly on the tangent bundle of the manifold. It will turn out that this is neither a limitation nor a drawback but the only natural choice for such coordinates. All geometric objects derived from a general connection on the tangent bundle, like the curvature or covariant derivatives, are objects living generically on the tangent bundle of the manifold and not on the manifold itself. Thus coordinates adapted to the geometry can only be special coordinates on the tangent bundle. The construction of  these coordinate systems will be based on the definition of the newly introduced tangent bundle exponential map which maps a double copy of the tangent space to the manifold to the tangent bundle. The locally autoparallel coordinates for general connections on the tangent bundle, respectively the Finslerian geodesic coordinates, will hopefully simplify proofs and calculations in the study of such non-metric geometries of manifolds in mathematics and physics.

We begin this article by a review of the concept of general connections on the tangent bundle in section \ref{sec:nctm}. Here we define the notation and lay all mathematical foundations needed throughout this article. In section \ref{sec:exp} we introduce the key concepts of this article: the new tangent bundle exponential map in Definition 3 and two locally autoparallel coordinate systems which have similar properties but differ in their interpretation in Definition 4 and Definition 5. Having studied the properties of the tangent bundle exponential map and the locally autoparallel coordinates in general we apply them to Finslerian geometries in section \ref{sec:finsexp}. Here the key results are Corollary 2 and Corollary 3, the expansions of the geometry defining Finsler Lagrangian in a power series in a neighborhood of a point on the manifold in the two new coordinate systems. Finally we conclude in section \ref{sec:disc} with a remark on a possible interpretation of the newly introduced coordinates from a physical point of view.

\section{General connections on the tangent bundle}\label{sec:nctm}
We recall the basics facts about general connections on the tangent bundle of a manifold needed to formulate the tangent bundle exponential map in section \ref{sec:exp}. We begin by introducing briefly the tangent bundle $TM$ of a manifold $M$ as fibre bundle, we recall manifold induced coordinates and the notion of the  vertical tangent space of $TM$. The latter then leads to the definition of general connections, the horizontal tangent space, the corresponding curvature and the associated Berwald linear connection and its autoparalllels. The autoparallels of the Berwald linear connection are the key ingredient in the definition of the tangent bundle exponential map in the next section. More details on these topics can be found for example in the textbooks \cite{NODG} or \cite{FLG}.

\subsection{The tangent bundle and its tangent spaces}
The tangent bundle $TM$ of an $n$-dimensional smooth manifold $M$ is the union of all tangent spaces $T_pM$ of $M$
\begin{equation}
TM=\bigcup_{p\in M} T_pM\,.
\end{equation}
It is itself a $2n$-dimensional manifold and carries naturally the structure of a fibre bundle $(TM, M, \pi, \mathbb{R}^n)$ with projection map $\pi$ which associates to each vector $Y\in T_pM\subset TM$ its base point $\pi(Y)=p$. On $TM$ there exists a special choice of coordinates, the so called manifold induced coordinates, which will be used as reference coordinates throughout this article. They are defined as follows. Consider coordinates $\{x\}$ in a neighborhood of a point $p\in M$. Then, a vector $Y$ in $T_pM$, which is a point in $TM$, can be expressed in the local coordinate basis of $T_pM$ as $Y=y^a\frac{\partial}{\partial x^a}_{|p}$. The manifold induced coordinates of the tangent bundle point $Y$ are the numbers $(x,y)$. From now on upper case letters $Y$ denote points in $TM$ while lower case letters $(x,y)$ denote the coordinate representation of the point. In these coordinates the projection map $\pi$ takes the form $\pi(x,y)=x$. From the definition of the coordinates it is clear that a coordinate change $x\mapsto \tilde x(x)$ on $M$ induces the following coordinate change on $TM$
\begin{equation}\label{eq:mfcoord}
(x,y)\mapsto(\tilde x(x), \tilde y(x,y)),\ \tilde y^a(x,y)=\frac{\partial \tilde x^a}{\partial x^q}(x)y^q=A^q{}_a(x)y^q\,.
\end{equation}
Moreover the manifold induced coordinates of $TM$ immediately lead to an induced coordinate basis of the tangent and cotantgent spaces $T_YTM$ and $T_Y^*TM$ of the tangent bundle denoted by ${\big\{\partial_a=\frac{\partial}{\partial x^a}, \bar\partial_a=\frac{\partial}{\partial y^a}\big\}}$ respectively $\{dx^a, dy^a\}$. They transform under a coordinate change on the manifold as
\begin{eqnarray}\label{eq:coordbs}
\big\{\partial_a,\ \bar\partial_a\big\}&=&\big\{A^q{}_a \tilde \partial_q+y^b \partial_a A^q{}_b \tilde {\bar\partial}_q,\ A^q{}_a \tilde {\bar \partial}_q \big\}\\
\big\{dx^a, dy^a\big\}&=&\big\{A^{-1a}{}_qd\tilde x^q,\  \tilde y^b\tilde \partial_q A^{-1a}{}_b d\tilde x^q + A^{-1a}{}_q d\tilde y^q \big\}\,.
\end{eqnarray}
The differential of the bundle projection $d\pi_{(x,y)}$ annihilates the $\bar\partial$ part of a vector $Z\in T_{(x,y)}TM$
\begin{equation}
d\pi_{(x,y)} (Z^a\partial_a{}_{|(x,y)}+\bar Z^a\bar\partial_a{}_{|(x,y)})=Z^a\partial_a{}_{|x}\,.
\end{equation}
The kernel of the projection $d\pi_{(x,y)}$ is called the vertical tangent space $V_{(x,y)}TM\subset T_{(x,y)}TM$ to $TM$. The existence of this canonical subspace of the tangent space of the tangent bundle leads to the notion of a general connection $\omega$: it is the projection of $T_{(x,y)}TM$ to $V_{(x,y)}TM$. Its kernel $H_{(x,y)}TM$ is called the horizontal tangent space and defines a complement to $V_{(x,y)}TM$ in $T_{(x,y)}TM$ such that $T_{(x,y)}TM=H_{(x,y)}TM\oplus V_{(x,y)}TM$.

\subsection{Connections, the horizontal tangent space, curvature and a covariant derivative}\label{sec:conn}
Let $\omega$ be a projection from $T_YTM$ to $V_YTM$. In manifold induced coordinates $(x,y)$ of the tangent bundle the projection $\omega$ is the following $(1,1)$-tensor on $TM$
\begin{equation}\label{eq:conn}
\omega_{(x,y)}=(dy^a+N^a{}_b(x,y)dx^b)\otimes \bar\partial_a\,.
\end{equation}
Such a projection is called connection and defined through its connection coefficients $N^a{}_b(x,y)$. In case these connection coefficients are linear in their $y$-dependence, so called linear connections, we can write $N^a{}_b(x,y)=\Gamma^a{}_{bc}(x)y^b$. The $\Gamma^a{}_{bc}(x)$ are the coefficients of an affine connection on $M$ which appears for example in Riemannian geometry. Other important special cases are homogeneous and symmetric connections which appear naturally in Finslerian geometries \cite{BCS, FLG}.  In this cases the connection coefficients are not linear but only homogeneous of degree one with respect to the $y$-coordinates $N^a{}_b(x,\lambda y)=\lambda N^a{}_b(x, y)$ and share the following symmetry property $\bar\partial_bN^a{}_c=\bar\partial_cN^a{}_b$. The second coordinate system we introduce in section \ref{sec:exp} will only be available for homogeneous and symmetric connections. Moreover we will meet them in section \ref{sec:finsexp} where we study the tangent bundle exponential map in a Finsler geometric setting. 

The complement of the vertical tangent space $V_{(x,y)}TM$ to $TM$ is given by the kernel of the connection and is called the horizontal tangent space $H_{(x,y)}TM$.  A connection on the tangent bundle gives rise to the horizontal-vertical basis of the tangent and cotangent spaces 
\begin{eqnarray}
T_{(x,y)}TM&=&H_{(x,y)}TM\oplus V_{(x,y)}TM\nonumber\\
&=&<\delta_a=\partial_a-N^b{}_a(x,y)\bar\partial_b>\oplus <\bar\partial_a>\\
T^*_{(x,y)}TM&=&(H_{(x,y)}TM)^*\oplus (V_{(x,y)}TM)^*\nonumber\\
&=&<dx^a>\oplus<\delta y^a=dy^a+N^a{}_b(x,y)dx^b>\,.
\end{eqnarray}
Due to the transformation behaviour of the nonlinear connection coefficients under a manifold induced coordinate transformation (\ref{eq:mfcoord})
\begin{equation}\label{eq:ntrafo}
\tilde N^a{}_b(\tilde x(x),\tilde y(x,y))=N^p{}_q(x,y)A^a{}_p A^{-1q}{}_b+\tilde y^q \partial_b A^{-1 p}{}_q A^a{}_p\,,
\end{equation}
since $\omega_{(x,y)}=(dy^a+N^a{}_b(x,y)dx^b)\otimes \bar\partial_a=(d\tilde y^a+\tilde N^a{}_b(\tilde x(x),\tilde y(x,y))d\tilde x^b)\otimes \tilde{\bar\partial}_a$, see appendix~\ref{app:nlintrafo}, the horizontal-vertical basis of $T_{(x,y)}TM$ and $T^*_{(x,y)}TM$ transforms nicely as 
\begin{equation}\label{eq:trafo}
\big\{\delta_a,\ \bar\partial_a\big\}=\big\{A^q{}_a \tilde\delta_q,\ A^q{}_a \bar\partial_q\big\},\ \big\{dx^a,\ \delta y^a\big\}=\big\{A^{-1a}{}_qd\tilde x^q,\ A^{-1a}{}_q \delta\tilde y^q\big\}\,.
\end{equation}
The curvature $R$ of the connection $\omega$ is defined as
\begin{equation}\label{eq:rcomp}
R=R^a{}_{bc}(x,y)dx^b\wedge dx^c\otimes \bar\partial_a=[\delta_b, \delta_c]^a\bar\partial_a=(\delta_c N^a{}_b(x,y)-\delta_b N^a{}_c(x,y))\bar\partial_a\,.
\end{equation}
It measures the integrability of the horizontal tangent spaces and becomes equivalent to the usual Riemann curvature tensor in case the connection coefficients are linear in their dependence on $y$. Under manifold induced coordinate transformations the curvature transforms as if it were a tensor on the base manifold, except that its components depend not only on the position on the manifold $x$ but also on the directions~$y$
\begin{equation}
R=R^a{}_{bc}(x, y)A^{-1b}{}_p A^{-1c}{}_q A^{d}{}_a d\tilde x^p\wedge d\tilde x^q\otimes \tilde{\bar\partial}_d=\tilde R^d{}_{pq}(\tilde x,\tilde y)d\tilde x^p\wedge d\tilde x^q\otimes \tilde{\bar\partial}_d\,.
\end{equation}
This transformation behaviour can be checked with help of the transformation formulae (\ref{eq:ntrafo}) and (\ref{eq:trafo}) and the definition of the components of $R$ in the horizontal-vertical basis in equation (\ref{eq:rcomp}). 

In order to construct the tangent bundle exponential map we need curves on $TM$ which reflect the geometry defined by a general connection, in other words its autoparallels.  Consider a general curve $\gamma(t)\in TM$. In manifold induced coordinates its tangent vector can be expressed in the coordinate basis (\ref{eq:coordbs}) respectively in the horizontal-vertical basis (\ref{eq:trafo}) of $TTM$
\begin{eqnarray}\label{eq:dotgamma}
\gamma(t)=(x(t),y(t)),\ \dot\gamma(t)&=&\dot x^a(t)\partial_a+\dot y^a(t) \bar\partial_a\\
&=&\dot x^a(t)\delta_a+(\dot y(t)+N^a{}_b(\gamma(t))\dot x^b(t))\bar\partial_a\,.
\end{eqnarray}

\vspace{6pt}\noindent\textbf{Definition 1.}\label{def:nlautopara0}
\textit{Let $N^a{}_b(x,y)$ be the connection coefficients of a connection $\omega$ on the tangent bundle of a manifold $M$. A curve $x_U(t)\in M$ is an autoparallel of the connection $\omega$ if its canonical lift to the tangent bundle $X_U(t)=(x_U(t),\dot x_U(t))$ satisfies
\begin{eqnarray}\label{eq:nlautopara0}
0&=&\omega(\dot X_U(t))=(\ddot x_{U}^a + N^a{}_b(x_{U},\dot x_{U})\dot x_U^b)\bar\partial_a\label{eq:horr0}\,,
\end{eqnarray}
with initial conditions $x_U(0)=x_0,\ \dot X_U(0)=U^{H_{(x_0,u)}}=u^a\delta_a\in H_{(x_0,u)}TM$, where $U^{H_{(x_0,u)}}$ is the horizontal lift of the vector $U=u^a\partial_a\in T_{x_0}M$ to $H_{(x_0,u)}TM$.}

\noindent As a matter of fact these autoparallels of the general connection alone do not carry enough information to define the tangent bundle exponential map. More general curves on $TM$ are needed among which the above autoparallels are a special subclass. The set of curves we are looking for are the horizontal autoparallels of the Berwald linear connection. The Berwald linear connection, associated to the nonlinear connection $\omega$, is defined via covariant derivatives on $TM$ acting on the horizontal-vertical basis of $T_{(x,y)}TM$ as follows
\begin{eqnarray}\label{eq:berwald}
\nabla^B_{\delta_b}\delta_c=D^a{}_{bc}(x,y)\delta_a,&\ & \nabla^B_{\delta_b}\bar\partial_c=D^a{}_{bc}(x,y)\bar\partial_a\\
\nabla^B_{\bar\partial_b}\delta_c=0,&\ & \nabla^B_{\bar\partial_b}\bar\partial_c=0\,.
\end{eqnarray}
with $D^a{}_{bc}(x,y)=\bar\partial_b N^a{}_{c}(x,y)$. This covariant derivative is purely defined from the nonlinear connection and respects the horizontal-vertical split structure of $T_YTM$. Its horizontal autoparallels, i.e. curves $\gamma(t)\in TM$ with horizontal tangent, $\omega(\dot\gamma)=0$, satisfying $\nabla^B_{\dot\gamma}\dot\gamma=0$ will serve as image of the exponential map and as coordinate lines. 

\vspace{6pt}\noindent\textbf{Definition 2.}\label{def:nlautopara}
\textit{Let $N^a{}_b(x,y)$ be the connection coefficients of a connection $\omega$ on the tangent bundle of a manifold $M$. A curve $\gamma_{(U,V)}(t)=(x_U(t), y_V(t))$ is a horizontal autoparallel of the Berwald linear connection induced by the nonlinear connection $\omega$ if it satisfies
\begin{eqnarray}
0&=&\omega(\dot\gamma_{(U,V)})=(\dot y_{V}^a + N^a{}_b(x_{U},y_{V})\dot x_u^b)\bar\partial_a\label{eq:horr}\\
0&=&\nabla^B_{\dot\gamma_{(U,V)}}\dot\gamma_{(U,V)}=(\ddot x_{U}^a + \bar\partial_cN^a{}_b(x_{U} , y_{V})\dot x_{U}^b \dot x_{U}^c)\delta_a\label{eq:horrauto}\,,
\end{eqnarray}
with initial position $\gamma_{(U,V)}(0)=V=(x_0,v)\in T_{x_0}M\subset TM$ and initial velocity $\dot\gamma_{(U,V)}(0)=U^{H_{(x_0,v)}}=u^a\delta_a\in H_{(x_0,v)}TM$, where $U^{H_{(x_0,v)}}$ is the horizontal lift of the vector $U=u^a\partial_a\in T_{x_0}M$ to $H_{(x_0,v)}TM$.}

\noindent These curves have the following important property:

\vspace{6pt}\noindent\textbf{Theorem 1.}
\textit{Let $\gamma_{(U,V)}(t)=(x_U(t), y_V(t))$ be a horizontal autoparallel of the Berwald linear connection, then
\begin{equation}
\gamma_{(\alpha U, V)}(1)=\gamma_{(U,V)}(\alpha)\,.
\end{equation}}

\vspace{6pt}\noindent\textit{Proof of Theorem 1.}
Define $\sigma(t)=\gamma_{(U,V)}(\alpha t)$ and calculate
\begin{equation}
\sigma(0)=(x_0, v),\ \dot\sigma(0)=\alpha u^a\delta_a\,.
\end{equation}
Thus $\sigma(t)$ is the curve $\gamma_{(\alpha U, V)}(t)$ and we can conclude for $t=1$ the statement of theorem 1. $\square$

\noindent For homogeneous and symmetric connections, i.e. connections with connection coefficients satisfying $N^a{}_b(x,\lambda y)=\lambda N^a{}_b(x,y)$ and $\bar\partial_cN^a{}_b=\bar\partial_bN^a{}_c$, distinguished horizontal autoparallels of the Berwald linear connection are the natural lifts of autoparallels of the connection to the tangent bundle.

\vspace{6pt}\noindent\textbf{Theorem 2.}
\textit{Let $N^a{}_b(x,y)$ be the connection coefficients of a homogeneous and symmetric connection $\omega$. The horizontal autoparallels $ \gamma_{(U,U)}(t)$ of the Berwald connection are natural lifts of the autoparallels of $\omega$ to the tangent bundle.}

\vspace{6pt}\noindent\textit{Proof of Theorem 2.}
The horizontal autoparallels $\gamma_{(U,U)}(t)=(x_U(t),y_U(t))$ of the Berwald linear connection satisfy $y^a_U(0)=u^a=\dot x_U^a(0)$ from the definition of the curves. A solution of the autoparallel equations (\ref{eq:horr}) and (\ref{eq:horrauto}) is $y_U(t)= \dot x_U(t)$ and $x_U(t)$ being an autoparallel of the general connection with initial conditions $x_U(0)=x_0$ and $\dot x_U(0)=U$. Thus by the uniqueness of the solutions of ordinary differential equations for given initial conditions: $ \gamma_{(U,U)}(t)$ is the natural lift $( x_U(t),\dot{ x}_U(t))$ of the autoparallel  $x_U(t)$ of the connection $\omega$. $\square$

\noindent In the context of Finslerian geometries in section \ref{sec:finsexp} Theorem 2 connects the horizontal autoparallels of the Berwald linear connection with the geodesics of the geometry.

The notions recapitulated during this section enable us to construct the tangent bundle exponential map for a general connection.

\section{The tangent bundle exponential map and locally autoparallel coordinates for general connections}\label{sec:exp}
In affine connection and metric geometry the exponential map identifies vectors in the tangent space to a manifold with points along curves adapted to the geometry of the manifold. In the previous section we considered a manifold whose geometry is based on a general connection on the tangent bundle with the result that the geometric objects like the covariant derivatives, connection coefficients and curvature are objects living on $TM$. Therefore the tangent bundle exponential map we construct here maps a double copy of the tangent space to the manifold to the tangent bundle. It will turn out that this map gives rise to coordinates on $TM$ in which the connection coefficients of a general connection vanish at the origin of the coordinate system. We begin with the definition of the exponential map and its properties before we use the map to introduce two kinds of locally autoparallel coordinates.

\subsection{The tangent bundle exponential map}\label{sec:tmexp}
We define the tangent bundle exponential map for general connections $\omega$ on $TM$ in a neighborhood of a point $p\in M\subset TM$ as map from $T_pM\times T_pM$ to $TM$.

\vspace{6pt}\noindent\textbf{Definition 3.}
\textit{Let $p$ be a point in $M$ with coordinates $x_0$, $\{\partial_a\}$ be the corresponding coordinate basis of $T_pM$ and let $\gamma_{(U,V)}(t)$ be a horizontal autoparallel of the Berwald linear connection associated to the general connection $\omega$ on the tangent bundle. The tangent bundle exponential map at $p$ is the mapping
\begin{eqnarray}
EXP_p:T_pM\times T_pM&\rightarrow& TM\\
(U, V)&\mapsto&\gamma_{(U,V)}(1)\,.
\end{eqnarray}}

\noindent In locally manifold induced coordinates $(x,y)$ it takes the form
\begin{equation}
EXP_p(u^a\partial_a, v^b\partial_b)=(E_x(u^a\partial_a, v^b\partial_b), E_y(u^a\partial_a, v^b\partial_b))=(x_U(1), y_V(1))\,,
\end{equation}
where $E_x$ and $E_y$ are the base point and the components of the vector $EXP_p(U, V)\in T_{x_U(1)}M\subset TM$. We now investigate the properties of this exponential map. First we analyse the smoothness properties of $EXP_p$.

\vspace{6pt}\noindent\textbf{Theorem 3.}
\textit{The exponential map $EXP_p$ is smooth on $T_pM\times T_pM\setminus A$ with
\begin{eqnarray}
A=&&\{(u^a\partial_a, v^b\partial_b) \in T_pM\times T_pM |\nonumber\\
&&N^a{}_b(x_0,v)u^b, \bar\partial_c N^a{}_b(x_0,v)u^b u^c \notin C^\infty(T_pM\otimes T_pM)\ \  \}\,.
\end{eqnarray}
The set A has to be discussed for each connection in consideration separately.}

\vspace{6pt}\noindent\textit{Proof of Theorem 3.} 
The point $(U,V)=(u^a\partial_a, v^b\partial_b)\in T_pM\times T_pM$ labels the initial data of the autoparallel $\gamma_{(U,V)}\in TM$. By the theory of ordinary differential equations everywhere where the autoparallel equations (\ref{eq:horr}) and (\ref{eq:horrauto}) are smooth, $\gamma_{(U,V)}$ depends smoothly on the initial data. Thus $EXP_p$ is smooth wherever $N^a{}_b(x_0,v)u^b$ and $\bar\partial_c N^a{}_b(x_0,v)u^b u^c$ are smooth in $C^\infty(T_pM\otimes T_pM)$. $\square$

In the following theorem we list properties of the tangent bundle exponential map needed to study the locally autoparallel coordinates which we define in the next section. 

\vspace{6pt}\noindent\textbf{Theorem 4.}
\textit{Properties of the exponential map:
\begin{enumerate}
	\item The exponential map $EXP_p$ maps curves $(t U, V)\in T_pM\times T_pM$ to horizontal autoparallels of the Berwald connection associated to the general connection $\omega$
	\begin{equation}
	EXP_p(t U, V)=\gamma_{(U,V)}(t)\,;
	\end{equation}
	\item For a homogeneous and symmetric connection $\omega$ the curves $(t U, U)\in T_pM\times T_pM$ are mapped to the natural lift $(x_U(t),\dot x_U(t))\in TM$ of autoparallels  $x_U(t)\in M$ of the connection;
	\item $EXP_p$ is the identity between $\{0\}\times T_pM$ and $T_pM$
	\begin{equation}
	EXP_p(0,V)=V;
	\end{equation}
	\item the first derivatives of $EXP_p$ at $(0,V)$with respect to the components $u^a$ and $v^a$ of $(U,V)\in T_pM\times T_pM$ are given by
	\begin{eqnarray}
	\frac{\partial}{\partial u^b}E_x^q(0,V)=\delta^q_b,\ && \frac{\partial}{\partial u^b}E_y^q(0,v)=-N^q{}_b(x_0, v);\\
	\frac{\partial}{\partial v^b}E_x^q(0,V)=0,\ && \frac{\partial}{\partial v^b}E_y^q(0,v)=\delta^q_b;
	\end{eqnarray}
	\item $EXP_p$ is a diffeomorphism around $(0, V)\in T_pM\times T_pM\setminus A$;
	\item the second derivative of $EXP_p$ with respect to the coordinate components $u^a$ of $U\in T_pM$ is given by
	\begin{eqnarray}
	\frac{\partial^2}{\partial u^b\partial u^c}E_x^q(0,V)&=&-\bar\partial_{(c}N^q{}_{b)}(x_0, v)\\
	\frac{\partial^2}{\partial u^b\partial u^c}E_y^q(0,V)&=&-\delta_{(c}N^a{}_{b)}(x_0,v)\nonumber\\
	&+&N^a{}_q(x_0,v)\bar\partial_{(b}N^q{}_{c)}(x_0, v)\,,
	\end{eqnarray}
	where the round brackets denote symmetrisation over the indices.
	\item the third derivative of $E_x$ with respect to the coordinate components $u^a$ of $U\in T_pM$ is given by
	\begin{eqnarray}
	\frac{\partial^3}{\partial u^b\partial u^c \partial u^d}E_x^q(0,V)&=&-\delta_{(d}\bar\partial_{c}N^q{}_{b)}(x_0, v)\nonumber\\
	&+&2\bar\partial_rN^q{}_{(b}(x_0, v)\bar\partial_cN^r{}_{d)}(x_0, v)\,.
	\end{eqnarray}
\end{enumerate}
}

\vspace{6pt}\noindent\textit{Proof of Theorem 4.} 
	\begin{enumerate}
	\item By definition of $EXP_p$ and the property of the horizontal autoparallels stated in Theorem 1
	\begin{equation}
	EXP_p(t u, v)=\gamma_{(t U,V)}(1)=\gamma_{(U,V)}(t)\,;
	\end{equation}
	\item by Theorem 2 of the previous section we know that $\gamma_{(U,U)}(t)$ is the the natural lift to $TM$ of the autoparallel $x_U(t)\in M$ of a homogeneous and symmetric connection $\omega$. Property $(1)$ of this theorem then shows
	\begin{equation}
	EXP_p(t u, u)=\gamma_{(t U,U)}(1)=\gamma_{(U,U)}(t)\,;
	\end{equation}
	\item by $(1)$ of this theorem
	\begin{equation}
	EXP_p(0,V)=\gamma_{(0,V)}(1)=\gamma_{(U,V)}(0)=(x_u(0),y_v(0))=(x_0,v)=V\,;
	\end{equation}
	\item differentiating $EXP_p(t U, V)=\gamma_{(t U,V)}(1)=\gamma_{(U,V)}(t)$ with respect to $t$ and evaluating at $t=0$ yields with help of equation~(\ref{eq:dotgamma}) or equation~(\ref{eq:horr})
	\begin{eqnarray}
	\frac{\partial}{\partial u^b}E_x^q(0,V)u^b \partial_q+\frac{\partial}{\partial u^b}E_y^q(0,V)u^b\bar\partial_q&=&\dot \gamma_{(U,V)}(0)\\
	&=&u^q \partial_q-N^q{}_b(x_0,v)u^b\bar\partial_q\,.\nonumber
	\end{eqnarray}
	To obtain the $v^a$ derivatives simply differentiate the following equation with respect to s at $s=1$
	\begin{equation}
	EXP_p(0,s V)=\gamma_{(0,s V)}(1)=\gamma_{(U,s V)}(0)=(x_0, s v)\,;
	\end{equation}
	\item by statement three of this theorem the determinant of the Jacobian of $EXP_p$ at $(0,V)$ is clearly non vanishing and exists for every $(0, V)\in T_pM\times T_pM\setminus A$. Thus by the implicit function theorem there exists a neighbourhood in $T_pM\times T_pM\setminus A$ around $(0,V)$ such that $EXP_p$ is a diffeomorphism;
	\item the second derivative of $E_x(t U,V)$ with respect to $t$ equals $\ddot x_U(t)$ and the second derivative of $E_y(t U,V)$ equals $\ddot y_V(t)$ and thus at $t=0$
	\begin{eqnarray}
	\frac{\partial^2}{\partial u^b\partial u^c}E_x^q(0,V)u^b u^c &=&\ddot x^q_{U}(0)=-\bar\partial_cN^q{}_b(x_0,v)u^bu^c\\
	\frac{\partial^2}{\partial u^b\partial u^c}E_x^q(0,V)u^b u^c &=&\ddot y^q_{V}(0)=-\delta_{c}N^q{}_{b}(x_0,v)\nonumber\\
	&+&N^q{}_a(x_0,v)\bar\partial_{b}N^a{}_{c}(x_0, v))u^bu^c\,,
	\end{eqnarray}
	where the $t$ derivative of equation (\ref{eq:horr}), and equation (\ref{eq:horrauto}) were used.
	\item the third derivative of $E_x(t U,V)$ with respect to $t$ equals $\dddot x_U(t)$ and thus at $t=0$
	\begin{eqnarray}
	&&\frac{\partial^3}{\partial u^b\partial u^c\partial u^d}E_x^q(0,V)u^b u^cu^d=\dddot x^q_{U}(0)\\
	&=&(-\delta_{(d}\bar\partial_{c}N^q{}_{b)}(x_0, v)+2\bar\partial_rN^q{}_{(b}(x_0, v)\bar\partial_cN^r{}_{d)}(x_0, v))u^b u^cu^d\nonumber\,.
	\end{eqnarray}
	where the $t$ derivative of equation (\ref{eq:horrauto}), and equation (\ref{eq:horr}) were used.~$\square$
\end{enumerate}
Observe that property five of Theorem 4 demonstrates explicitly that the second derivative of $EXP_p(0,V)$ with respect to $u$ is smooth wherever $\partial_cN^q{}_b(x_0,v)$ is smooth. In previous attempts to formulate an exponential map for homogeneous connections in a Finslerian geometry setting as a map from a tangent space $T_pM$ to $M$ this fails and led to the theorems on the non-existence of smooth normal coordinates in Finsler geometry \cite{Busemann, Rund}. In section \ref{sec:finsexp} we will investigate the tangent bundle exponential map in a Finsler geometric setting. Now in the next section we will present coordinates induced by the tangent bundle exponential map we defined here. 

\subsection{Locally autoparallel coordinates}\label{sec:lacs}
The exponential map introduced in the previous section allows us to define coordinates on $TM$ around all points $(x_0, y)\in T_{x_0}M\subset TM$, such that the nonlinear connection coefficients vanish $\tilde N^a{}_b(x_0,y)$ in $T_{x_0}M$. Here we introduce two such coordinate systems. It turns out that the first one, the extended locally autoparallel coordinates are such that the horizontal autoparallels of the Berwald linear connection associated to the general connection become straight lines. In case the general connection is homogeneous and symmetric also the autoparallels of the connection itself become straight lines since in this case they coincide with special autoparallels of the Berwald linear connection. However the extended locally autoparallel coordinate system has the disadvantage that it does not reduce to the usual normal coordinates known from the geometry of affine connections in case the connection coefficients are linear $N^a{}_b(x,y)=\Gamma^a{}_{bc}(x)y^c$. The cause of this feature is that we use the full freedom of the tangent bundle to construct the coordinates. To obtain coordinates which mimic the behaviour of the usual normal coordinates we introduce a second coordinate system, the standard locally autoparallel coordinates. They can only be constructed for homogeneous and symmetric connections on the tangent bundle and we will see that they are identical to the Douglas-Thomas normal coordinates when we apply these standard locally autoparallel coordinates to Finslerian geometries in the next section.

\vspace{6pt}\noindent\textbf{Definition 4.}
\textit{Let $\omega$ be a general connection on the tangent bundle and $p$ be a point in $M$ with manifold induced coordinates $x_0$. Let $(x, y)$ be manifold induced coordinates around $T_pM$ and express $(\tilde X,\tilde Y)\in T_pM\times T_pM\setminus A$ as $(\tilde X,\tilde Y)=(\tilde x^a\partial_a,\tilde y^a\partial_a)$ where $\{\partial_a\}$ is the coordinate basis of $T_pM$. Define new coordinates $(\tilde x, \tilde y)$ around $T_pM\subset TM$ via
\begin{eqnarray}\label{eq:lapc}
(x(\tilde x, \tilde y), y(\tilde x, \tilde y))=EXP_p(\tilde X, \tilde Y)=EXP_p(\tilde x^a\partial_a, \tilde y^a\partial_a)=\gamma_{(\tilde X, \tilde Y)}(1)\,.
\end{eqnarray}
The coordinates $(\tilde x, \tilde y)$ are called extended locally autoparallel coordinates.}

\noindent Here we used the coordinate basis components of the vectors $\tilde X$ and $\tilde Y$ to define the new coordinates. Equally well one could use the components of the vectors with respect to any other basis of $T_pM$ as coordinates. Before we investigate the properties of the locally autoparallel coordinates we demonstrate that they are well defined, i.e. that they are smooth and invertible.
 
From the definition of the coordinates $(\tilde x, \tilde y)$ it follows that the point $(x(0, \tilde y), y(0, \tilde y))=(x_0, \tilde y)$, i.e $(0,\tilde Y)\in T_pM\times T_pM$ gets mapped to $\tilde Y\in T_pM$. To study the behaviour and the properties of the new coordinates around $p$ we expand the defining coordinate transformation, equation (\ref{eq:lapc}), in a power series around $\tilde x=0$. Using the results from Theorem 4 we obtain
\begin{eqnarray}
x^q(\tilde x, \tilde y)
&=&x^q_0+\tilde x^q-\frac{1}{2} \bar\partial_cN^q{}_b(x_0, \tilde y)\tilde x^b\tilde x^c\nonumber\\
&-&\frac{1}{3!}(\delta_{d}\bar\partial_{c}N^q{}_{b}(x_0, \tilde y)-2\bar\partial_rN^q{}_{b}(x_0, \tilde y)\bar\partial_cN^r{}_{d}(x_0, \tilde y))\tilde x^b\tilde x^c\tilde x^d\\
&+&\mathcal{O}(\tilde x^4)\label{eq:trafo1}\nonumber\\
y^q(\tilde x, \tilde y)
&=&\tilde y^q-N^q{}_b(x_0,\tilde y)\tilde x^b\nonumber\\
&-&\frac{1}{2} (\delta_{c}N^q{}_{b}(x_0,\tilde y)-N^q{}_a(x_0,\tilde y)\bar\partial_{b}N^a{}_{c}(x_0, \tilde y))\tilde x^b\tilde x^c\\
&+&\mathcal{O}(\tilde x^3)\label{eq:trafo2}.\nonumber
\end{eqnarray}
The corresponding coordinate transformation matrix on $TM$ is given by
\begin{eqnarray}
\frac{\partial x^q}{\partial \tilde x^b}(\tilde x, \tilde y)&=&\delta^q_b-\bar\partial_{(c}N^q{}_{b)}(x_0, \tilde y)\tilde x^c\nonumber\\
&-&(\delta_{(d}\bar\partial_{c}N^q{}_{b)}(x_0, \tilde y)-2\bar\partial_rN^q{}_{(b}(x_0, \tilde y)\bar\partial_cN^r{}_{d)}(x_0, \tilde y))\tilde x^c\tilde x^d\\
&+&\mathcal{O}(\tilde x^3)\label{eq:cm1}\nonumber\\
\frac{\partial x^q}{\partial \tilde y^b}(\tilde x, \tilde y)&=&-\frac{1}{2} \bar\partial_b\bar\partial_cN^q{}_d(x_0, \tilde y)\tilde x^c\tilde x^d\nonumber\\
&-&\frac{1}{3!}\bar\partial_b(\delta_{d}\bar\partial_{c}N^q{}_{e}(x_0, \tilde y)-2\bar\partial_rN^q{}_{e}(x_0, v)\bar\partial_cN^r{}_{d}(x_0, \tilde y))\tilde x^e\tilde x^c\tilde x^d\\
&+&\mathcal{O}(\tilde x^4)\label{eq:cm2}\nonumber\\
\frac{\partial y^q}{\partial \tilde x^b}(\tilde x, \tilde y)&=&-N^q{}_b(x_0,\tilde y)\nonumber\\
&-&(\delta_{(c}N^a{}_{b)}(x_0,\tilde y)- N^q{}_a(x_0,\tilde y)\bar\partial_{(b}N^a{}_{c)}(x_0, \tilde y))\tilde x^c+\mathcal{O}(\tilde x^2)\label{eq:cm3}\\
\frac{\partial y^q}{\partial \tilde y^b}(\tilde x, \tilde y)&=&\delta^q_b-\bar\partial_bN^q{}_c(x_0,\tilde y)\tilde x^c\nonumber\\
&-&\frac{1}{2}\bar\partial_b(\delta_{c}N^q{}_{d}(x_0,\tilde y)- N^q{}_a(x_0,\tilde y)\bar\partial_{d}N^a{}_{c}(x_0, \tilde y))\tilde x^c\tilde x^d\\
&+&\mathcal{O}(\tilde x^3)\label{eq:cm4}\,.\nonumber
\end{eqnarray}
These indeed are well defined and invertible in a neighbourhood around $\tilde x=0$ since the determinant of the coordinate transformation matrix is nonvanishing 
\begin{equation}
\det(\delta^A_B+M^A{}_{Bc} \tilde x^c+ \mathcal{O}(\tilde x^2))=1-\tilde x^b[\bar\partial_{(b}N^q{}_{q)}(x_0, \tilde y)+\bar\partial_qN^q{}_b(x_0,\tilde y)]+\mathcal{O}(\tilde x^2)\,.
\end{equation}
Here $\delta^A_B$ denotes the eight dimensional unit matrix and $M^A{}_{Bc}$ the first order contributions in the equations (\ref{eq:cm1}) to (\ref{eq:cm4}). The inverse coordinate transformation as a power series expansion in $\Delta^a=x^a-x_0^a$ is given by
\begin{eqnarray}
\tilde x^q(x, y)&=&\Delta^q +\frac{1}{2} \bar\partial_cN^q{}_b(x_0, y)\Delta ^b\Delta^c+\mathcal{O}(\Delta^3)\label{eq:trafo3}\\
\tilde y^q(x, y)&=&y^q+N^q{}_b(x_0, y)\Delta^b\\
&+&\frac{1}{2} (\delta_{c}N^q{}_{b}(x_0,y)+2 N^r{}_c(x_0,y)\bar\partial_bN^q{}_r(x_0,y))\Delta^b\Delta^c\nonumber+\mathcal{O}(\Delta^3).\label{eq:trafo4}
\end{eqnarray}
We restrict ourselves here to a second order expansion in $\Delta^a$ since this suffices for all further aims of this article, especially for the study of the exponential map and the new coordinates in a Finslerian geometry. The inverse coordinate transformation matrix can be derived explicitly in a power series in $\Delta^a$ from the equations~(\ref{eq:trafo3}) and~(\ref{eq:trafo4})
\begin{eqnarray}
\frac{\partial \tilde x^q}{\partial x^b}(x, y)&=&\delta^q_b+\bar\partial_{(c}N^q{}_{b)}(x_0, y)\Delta^c+\mathcal{O}(\Delta^2)\label{eq:cm5}\\
\frac{\partial \tilde x^q}{\partial y^b}(x, y)&=&\frac{1}{2} \bar\partial_b\bar\partial_cN^q{}_d(x_0, y)\Delta^c\Delta^d+\mathcal{O}(\Delta^3)\label{eq:cm6}\\
\frac{\partial \tilde y^q}{\partial x^b}(x, y)&=& 
N^q{}_b(x_0, y)+(\delta_{(c}N^a{}_{b)}(x_0,y)\nonumber\\
&+&2N^r{}_{(c}(x_0,y)\bar\partial_{b)}N^q{}_{r}(x_0,y))\Delta^c+\mathcal{O}(\Delta^2)\label{eq:cm7}\\
\frac{\partial \tilde y^q}{\partial y^b}(x,y) &=&\delta^q_b+\bar\partial_bN^q{}_c(x_0,y)\Delta^c+\frac{1}{2}\bar\partial_b(\delta_{c}N^a{}_{b}(x_0,y)\nonumber\\
&+&2N^r{}_c(x_0,y)\bar\partial_dN^q{}_r(x_0,y))\Delta^c\Delta^d+\mathcal{O}(\Delta^3)\label{eq:cm8}\,.
\end{eqnarray}
Observe that this coordinate transformation on $TM$ is not a manifold induced one since $x\neq x(\tilde x)$ and $y^a\neq \tilde\partial_q x^a(\tilde x) \tilde y^q$. Even for a linear connection $N^a{}_b(x,y)=\Gamma^a{}_{bc}(x)y^c$ where $x(\tilde x, \tilde y)=x(\tilde x)$ the above coordinate transformation uses the freedom of the tangent bundle since still $ y^a\neq \tilde\partial_q x^a(\tilde x) \tilde y^q$ from third order expansion of $x$ in $\tilde x$ and second order expansion of $y$ in $\tilde x$ on. Thus the coordinate change in the direction coordinates $y$ is defined independently of the coordinate change in the position coordinates $x$ as it is the case for autoparallel coordinates known from the geometry of affine connections. The coordinate change considered here can never be induced by a coordinate change only on the manifold. It is always a more general coordinate change on the tangent bundle.

However it is possible to introduce another locally autoparallel coordinate system which is more restricted. These are coordinates on the tangent bundle for which the coordinate change in the direction coordinates $y$ is derived from the coordinate change in the position coordinates $x$ in a way that the behaviour of a manifold induced coordinate change is mimicked.

\vspace{6pt}\noindent\textbf{Definition 5.}
\textit{Let $p$ be a point in $M$ with manifold induced coordinates $x_0$. Let $(x, y)$ be manifold induced coordinates around $T_pM$ and express$(\tilde X,\tilde Y)\in T_pM\times T_pM\setminus A$ as $(\tilde X,\tilde Y)=(\tilde x^a\partial_a,\tilde y^a\partial_a)$ where $\{\partial_a\}$ is the coordinate basis of $T_pM$. Moreover let $\omega$ be a symmetric and homogeneous connection, i.e. its connection coefficients satisfy $\bar\partial_bN^a{}_c(x,y)=\bar\partial_cN^a{}_b(x,y)$ and $N^a{}_b(x,\lambda y)=\lambda N^a{}_b(x,y)$. Define new coordinates $(\tilde x, \tilde y)$ around $T_pM\subset TM$ via
\begin{eqnarray}\label{eq:lapc2}
x^q(\tilde x, \tilde y)=\pi(EXP_p(\tilde x^a\partial_a, \tilde y^a\partial_a))^q,\quad y^q(\tilde x, \tilde y)=\frac{\partial x^q}{\partial \tilde x^p}(\tilde x, \tilde y)\tilde y^p\,.
\end{eqnarray}
The coordinates $(\tilde x, \tilde y)$ are called standard locally autoparallel coordinates.}

\noindent Expanding the standard locally autoparallel coordinates in a power series with respect to $\tilde x$ yields the same expansion for $x(\tilde x, \tilde y)$ as displayed in equation (\ref{eq:trafo1}) for its extended locally autoparallel counterpart but differ after the first order in equation~(\ref{eq:trafo2}) in the $y(\tilde x, \tilde y)$ coordinates
\begin{eqnarray}
y^q(\tilde x, \tilde y)
&=&\tilde y^q-\bar\partial_{(c}N^q{}_{b)}(x_0,\tilde y)\tilde x^b\tilde y^b\nonumber\\
&-&(\delta_{(d}\bar\partial_cN^q{}_{b)}(x_0,\tilde y)-2\bar\partial_rN^q{}_{(d}(x_0,\tilde y)\bar\partial_{b}N^r{}_{c)}(x_0, \tilde y))\tilde x^b\tilde x^c\tilde y^d\\
&+&\mathcal{O}(\tilde x^3)\nonumber\\
&=&\tilde y^q-N^q{}_{b}(x_0,\tilde y)\tilde x^b\nonumber\\
&-&(\delta_{(d}\bar\partial_cN^q{}_{b)}(x_0,\tilde y)-2\bar\partial_rN^q{}_{(d}(x_0,\tilde y)\bar\partial_{b}N^r{}_{c)}(x_0, \tilde y))\tilde x^b\tilde x^c\tilde y^d\\
&+&\mathcal{O}(\tilde x^3)\label{eq:trafo5}\,.\nonumber
\end{eqnarray}
Due to its importance we stress one more time that the difference to the extended locally autoparallel coordinates appears only from second order in $\tilde x$ on due to the homogeneity and symmetry properties of the connection coefficients. For general connection coefficients, i.e. non-homogeneous and non-symmetric ones, the extended and standard locally autoparallel coordinates would differ at first order in $\tilde x$ and the standard locally autoparallel coordinates would not deserve their name, as we will see in Theorem~6 below. However the  standard locally autoparallel coordinates have one advantage over the extended locally autoparallel coordinates, they mimic the behaviour of manifold induced coordinates due to the definition of the $y$ coordinate transformation from the $x$ coordinate transformation, they are the Douglas-Thomas normal coordinates in the application to Finslerian geometries, as we will discuss in section \ref{ssec:expFins}, and reduce to the usual normal coordinates known from Riemannian geometry in case of linear connection coefficients $N^a{}_{b}(x,y)=\Gamma^a{}_{bc}(x)y^c$.

Finally we prove the name giving properties of both locally autoparallel coordinate systems, summarised in the following theorems. The proof of these properties relies on the explicit expression of the first order expansion of the manifold induced coordinates $(x,y)$ in $\tilde x$.

\vspace{6pt}\noindent\textbf{Theorem 5.}
\textit{Let $\{\delta_a, \bar\partial_a\}$ and $\{dx^a,\delta y^a\}$ be the horizontal-vertical basis of $T_YTM$ and $T^*_YTM$ induced by manifold induced coordinates $(x,y)$ around $T_pM\subset TM$ with $x_0$ being the coordinates of $p$. They are identical to the coordinate basis $\{\tilde \partial_a, \tilde{\bar\partial}_a\}$ and $\{d\tilde x^a,d \tilde y^a\}$ induced by both locally autoparallel coordinates $(\tilde x, \tilde y)$, with $Y=(x_0, y)$ in manifold induced coordinates and $Y=(0,y)$ in the locally autoparallel coordinates.}

\vspace{6pt}\noindent\textit{Proof of Theorem 5.}
Expand the horizontal-vertical basis vector fields around $Y$ into the manifold induced coordinate basis and expand these in the coordinate basis of the locally autoparallel coordinates
\begin{eqnarray}
\bar\partial_a{}_{|(x,y)}&=&\bar\partial_a\tilde x^p\tilde\partial_p+\bar\partial_a\tilde y^q\tilde{\bar\partial}_q\\
\delta_a{}_{|(x,y)}&=&\partial_a\tilde x^p\tilde\partial_p+\partial_a\tilde y^q\tilde{\bar\partial}_q-N^b{}_a(x,y)(\bar\partial_b\tilde x^p\tilde\partial_p+\bar\partial_b\tilde y^q\tilde{\bar\partial}_q)\\
dx^a{}_{|(x,y)}&=&\tilde\partial_p x^a d\tilde x^p+\tilde{\bar\partial}_qx^a d\tilde y^q\\
\delta y^a{}_{|(x,y)}&=&\tilde\partial_p y^a d\tilde x^p+\tilde{\bar\partial}_q y^a d\tilde y^q+N^a{}_b(x,y)(\tilde\partial_p x^a d\tilde x^p+\tilde{\bar\partial}_qx^a d\tilde y^q)\,.
\end{eqnarray}
Evaluating these equations at $Y=(x_0, y)\in T_pM$ with help of the coordinate transformations (\ref{eq:cm1}) to (\ref{eq:cm4}) and (\ref{eq:cm5}) to (\ref{eq:cm8}), which are identical for both locally autoparallel coordinate systems to first order, yields the desired result
\begin{eqnarray}
\bar\partial_a{}_{|(x_0,y)}&=&\tilde{\bar\partial}_a,\ \delta_a{}_{|(x_0,y)}=\tilde\partial_a,\  dx^a{}_{|(x_0,y)}= d\tilde x^a,\ \delta y^a{}_{|(x_0,y)}= d\tilde y^a\,.\square
\end{eqnarray}
In this sense the locally autoparallel coordinates disentangle fibre directions and manifold directions in $T_YTM$. The most important feature of the locally autoparallel coordinates is formulated in the following theorem.

\vspace{6pt}\noindent\textbf{Theorem 6.}
\textit{Let $(\tilde x, \tilde y)$ be extended or standard locally autoparallel autoparallel coordinates around $Y\in T_pM\subset TM$. The extended or standard locally autoparallel coordinates of $Y\in T_pM$ are $Y=(0,y)$ and the manifold induced coordinates are $Y=(x_0,y)$. Then the nonlinear connection coefficients $\tilde N^a{}_b(0,y)$ vanish.}

\vspace{6pt}\noindent\textit{Proof of Theorem 6.}
The coordinate transformation matrices between the manifold induced $(x,y)$ coordinates and both locally autoparallel coordinates $(\tilde x, \tilde y)$ at $Y=(x_0, y)$ can be read of from equations (\ref{eq:cm1}) to (\ref{eq:cm4}) and (\ref{eq:cm5}) to (\ref{eq:cm8}) to be
\begin{eqnarray}
&&\frac{\partial x^q}{\partial \tilde x^b}(0, y)=\delta^q_b,\hspace{60pt} \frac{\partial x^q}{\partial \tilde y^b}(0, y)=0,\\
&& \frac{\partial y^q}{\partial \tilde x^b}(0, y)=-N^q{}_b(x_0,\tilde y),\hspace{13pt} \frac{\partial y^q}{\partial \tilde y^b}(0, y) =\delta^q_b,\\
&&\frac{\partial \tilde x^q}{\partial x^b}(x_0, y)=\delta^q_b,\hspace{55pt} \frac{\partial \tilde x^q}{\partial y^b}(x_0, y)=0,\\
&& \frac{\partial \tilde y^q}{\partial x^b}(x_0, y)=N^q{}_b(x_0, y),\hspace{17pt} \frac{\partial \tilde y^q}{\partial y^b}(x_0, y) =\delta^q_b\,.
\end{eqnarray}
It is identical for both kinds of locally autoparallel coordinates introduced. From the transformation behaviour of the connection $\omega$ under a general coordinate transformation, displayed in appendix \ref{app:nlintrafo}, it follows immediately that the nonlinear connection coefficients at $(0,\tilde y)$ transform similarly like under a manifold induced coordinate transformation as calculated in equation (\ref{eq:ntrafo}). The result is
\begin{eqnarray}\label{eq:tildenvanish}
\tilde N^a{}_b(0, y)&=&\frac{\partial y^p}{\partial \tilde x^b}(0, y)\frac{\partial \tilde y^a}{\partial y^p}(x(0, y),y(0,  y))\nonumber\\
&+&N^p{}_i(x(0, y),y(0,  y))\frac{\partial x^i}{\partial \tilde x^b}(0, y)\frac{\partial\tilde  y^a}{\partial y^p}(x(0, y),y(0,  y))\nonumber\\
&=&N^a{}_b(x_0,  y)-N^a{}_b(x_0, y)=0\,.\ \square
\end{eqnarray}
With this proof of Theorem 6 we have clearly demonstrated that for every $p\in M$ there exit coordinates around $T_pM\subset TM$ such that in these coordinates the connection coefficients vanish for all $Y\in T_pM\subset TM$. Moreover by the definition of the coordinates and the first two statements of Theorem 4 in the previous section: straight lines in the $\tilde x$ coordinates $(t \tilde x, \tilde y)$ are horizontal autoparallels of the Berwald connection, and for homogeneous and symmetric connections special straight lines $(t \tilde x, \tilde x)$ are even the autoparallels of the connection $\omega$.

In the next section we apply this very general construction to Finsler geometry. There the geodesic equation of the manifold gives rise to a unique homogeneous and symmetric connection and it turns out that the exponential map maps straight lines $(t U,U)\in T_pM\times T_pM$ smoothly to these geodesics.

\section{The Exponential map and locally autoparallel coordinates for Finslerian geometries}\label{sec:finsexp}

Finsler geometry is a long known and straightforward generalisation of metric geometry. It is the natural stage on which a more general connection than a linear connection appears to describe the geometry of a manifold, namely the so called Cartan nonlinear connection. Its connection coefficients are homogeneous of degree one with respect to the $y$ coordinates, symmetric and it generalises the well known Levi-Civita connection from metric geometry. Here we apply the construction of the tangent bundle exponential map and the locally autoparallel coordinates from the previous section. After a short introduction to Finslerian geometries we show that the tangent bundle exponential map identifies straight lines with Finsler geodesics and we expand the geometry defining object, the Finsler Lagrangian, in a power series with respect to the new $\tilde x$ coordinates. The main results, summarised in Corollary~2 and Corollary~3, are that the Finsler Lagrangian is independent of $\tilde x$ at least up to cubic order in extended locally autoparallel coordinates and  depended on $\tilde x$ only from quadratic order on in standard locally autoparallel coordinates. Finally we present the components of the curvature of the Cartan nonlinear connection in standard locally autoparallel coordinates.

\subsection{Finslerian geometries and Finsler spacetimes}
A Finslerian geometric manifold is a manifold $M$ equipped with a length measure $S$ for curves $\gamma$ on $M$ 
\begin{equation}\label{eq:flength}
S[\gamma]=\int d\tau F(\gamma, \dot\gamma).
\end{equation}
The length measure is defined through function $F(x,y)$ on $TM$, called the Finsler function, which is homogeneous of degree one with respect to the $y$-coordinates. The original development of Finsler geometry goes back to 1918 \cite{finsler}. Since then Finsler geometry has been well developed, see for example the book \cite{BCS}, and works fine as generalisation of Riemannian metric geometry as long as one considers only Finsler functions whose hessian, the so called Finsler metric, 
\begin{equation}
g^F_{ab}(x,y)=\frac{1}{2}\bar\partial_a\bar\partial_b F^2(x,y)
\end{equation}
is positive definite. When it comes to the application in physics one tries to formulate a generalisation of Lorentzian metric geometry in terms of a Finsler function. Here immediately issues on the well definedness of the geometry formulated in terms of $F$ appear. Due to the existence of non-trivial null directions $F^2$ is not differentiable everywhere and thus the geometry of the manifold does not exist wherever $F^2$ is not smooth \cite{Pfeifer:2011tk}. In order to overcome these issues many proposals have been made like the exclusion of the geometry on the nontrivial null structure by Asanov \cite{Asanov} or the restriction to specific classes of indefinite Finsler spaces by Beem \cite{Beem}. In previous works \cite{Pfeifer:2011xi, Pfeifer:2011tk} we have generalised the definition of indefinite Finsler spaces from Beem to include larger classes of Finslerian geometries without running into issues on the existence of the geometry along non-trivial null directions. Our definition of a Finsler spacetime is the following:

\vspace{6pt}\noindent\textbf{Definition 6.}
\textit{A Finsler spacetime $(M,L)$ is a four dimensional, connected, Hausdorff, paracompact, smooth manifold~$M$ equipped with a continuous function, called the Finsler Lagrangian, $L:TM\rightarrow\mathbb{R}$ on the tangent bundle which has the following properties:
\begin{enumerate}[(i)]
\item $L$ is smooth on the tangent bundle without the zero section $TM\setminus\{(x,0)\}$;
\item $L$ is positively homogeneous of real degree $r \ge 2$ with respect to the fibre coordinates of $TM$,
\begin{equation}\label{eqn:hom}
L(x,\lambda y)  = \lambda^r L(x,y) \quad \forall \lambda>0\,;
\end{equation}
\item \vspace{-6pt}$L$ is reversible in the sense 
\begin{equation}\label{eqn:rev} 
|L(x,-y)|=|L(x,y)|\,;
\end{equation}
\item \vspace{-6pt}the Hessian $g^L_{ab}$ of $L$ with respect to the fibre coordinates $y$ is non-degenerate on $TM\setminus B$ where~$B$ has measure zero and does not contain the null set $\{(x,y)\in TM\setminus\{(x,0)\}\,|\,L(x,y)=0\}$,
\begin{equation}
g^L_{ab}(x,y) = \frac{1}{2}\bar\partial_a\bar\partial_b L\,;
\end{equation}
\item \vspace{-6pt}the unit timelike condition holds, i.e., for all $x\in M$ the set 
\begin{eqnarray}
\Omega_x&=&\Big\{y\in T_xM\,\Big|\, |L(x,y)|=1,\nonumber\\
&&\ g^L_{ab}(x,y)\textrm{ has signature }(\epsilon,-\epsilon,-\epsilon,-\epsilon)\,,\, \epsilon=\frac{|L(x,y)|}{L(x,y)}\Big\}
\end{eqnarray}
contains a non-empty closed connected component $S_x\subset \Omega_x\subset T_xM$.
\end{enumerate}
The Finsler function associated to $L$ is $F(x,y) = |L(x,y)|^{1/r}$ and the Finsler metric $g^F_{ab}=\frac{1}{2}\bar\partial_a \bar\partial_b F^2$.}

For all details on the necessity of the requirements in our definition of Finsler spacetimes for physics see \cite{Pfeifer:2011xi,Pfeifer:2011tk}. Observe that for homogeneity $r=2$ and the degeneracy set $B=\emptyset$ Finsler spacetimes are Beem's indefinite Finsler spaces and without the assumption $(v)$ on the signature of the L-metric and the degeneracy set $B=\emptyset$ Finsler spacetimes contain the positive definite Finsler spaces. These assumptions will not play a role in the analysis of the exponential map on Finsler spacetimes, thus the results obtained here carry over to these other Finsler geometric settings.

The geometry of a Finsler spacetime $(M,L)$ is determined by the unique Cartan nonlinear connection on the tangent bundle of the manifold $M$, which is a special case of the situation discussed in section \ref{sec:nctm}.

\vspace{6pt}\noindent\textbf{Definition 7.}\label{def:cnlfs}
\textit{Let $(M,L)$ be a Finsler spacetime. The unique Cartan nonlinear connection on the tangent bundle is defined by the connection coefficients in manifold induced coordinates
\begin{equation}\label{eq:FSnonlin}
N^a{}_{b}(x,y)=\frac{1}{4}\bar\partial_b \Big[g^{Laq}(x,y)\big(y^p\partial_p\bar\partial_qL(x,y)-\partial_qL(x,y)\big)\Big]\,.
\end{equation}}
The connection coefficients of the Cartan nonlinear connection are one homogeneous with respect to the $y$-coordinates. It defines the geometry of the Finsler spacetime as described in section \ref{sec:conn}: the splitting of $T_YTM$ in vertical and horizontal tangent space $V_YTM$ and  $H_YTM$, the curvature $R$ and covariant derivatives. A consequence of the precise form of the nonlinear connection coefficients is that the Finsler Lagrangian is horizontally constant
\begin{equation}
\delta_a L(x,y)=0\,,
\end{equation}
and that the connection coefficients are symmetric in the sense
\begin{equation}\label{eq:finsgeod}
\bar\partial_bN^a{}_c(x,y)=\bar\partial_cN^a{}_b(x,y)\,.
\end{equation}
Geodesics on a Finsler spacetimes, i.e. curves $\gamma$ on $M$ which extremise the length functional (\ref{eq:flength}), are solutions to the geodesic equation
\begin{equation}
\ddot x^a+N^a{}_b(x,\dot x)\dot x^b=0\,.
\end{equation}

We now study the tangent bundle exponential map defined in section \ref{sec:tmexp} on Finsler spacetimes. It turns out that special horizontal autoparallels of the Berwald connection are Finslerian geodesics and thus we find that straight lines $(t U, U)\in T_pM\times T_pM$ are smoothly mapped to Finsler geodesics. Afterwards we derive the locally autoparallel coordinates introduced in section \ref{sec:lacs} for the Cartan nonlinear connection and calculate the components of the the curvature of the connection in these coordinates.

\subsection{The exponential map and Finsler geodesics}\label{ssec:expFins}
The horizontal autoprallels $\gamma_{(U,V)}(t)=(x_U(t),y_V(t))\in TM$ of the Berwald linear connection associated to the Cartan nonlinear connection satisfy the equations 
\begin{eqnarray}\label{eq:finsberwauto}
0&=&\dot y_{V}^a + N^a{}_b(x_{U},y_{V})\dot x_U^b=\dot y_{V}^a + \bar\partial_c N^a{}_b(x_{U},y_{V}) y^c_V\dot x_U^b\label{eq:fhorr}\\
0&=&\ddot x_{U}^a + \bar\partial_cN^a{}_b(x_{U} , y_{V})\dot x_{U}^b \dot x_{U}^c\label{eq:fauto}\,,
\end{eqnarray}
as discussed in Definition 2. Now observe that for $y_v(t)=\dot x_u(t)$ the autoparallel equations reduce to the Finsler geodesic equation (\ref{eq:finsgeod}) due to the homogeneity and symmetry of the connection coefficients $\bar\partial_cN^a{}_b(x,y) y^c=\bar\partial_bN^a{}_c(x,y) y^c=N^a{}_b(x,y)$. We like to remark here that the equations (\ref{eq:finsberwauto}), which are the fundamental ingredient in the definition of the autoparallel coordinates in this paper are identical to the equations (8-10) in \cite{Douglas1927}. There Douglas explains that Thomas pointed him to the fact that these equations lead to the normal coordinates he constructed throughout his article. Here we use the equations to construct two coordinate systems. The first system are the extended locally autoparallel coordinates, which use the full freedom of the tangent bundle, and the second system are the standard locally autoparallel coordinates which are more restricted and coincide with the Douglas-Thomas normal coordinates. In the next section we expand the Finsler Lagrangian of a Finsler spacetime in both coordinate systems.

The properties of the Cartan nonlinear connection coefficients ensure that the exponential map is smooth nearly everywhere. The set $A\subset T_pM\times T_pM$ which has to be omited, where the connection coefficients are not smooth, is the zero vector of $T_pM$ and the possible non empty set $B$ in the definition of Finsler spacetimes where the $L$ metric is not invertible.

\vspace{6pt}\noindent\textbf{Theorem 7.}
\textit{Let $(M,L)$ be a Finsler spacetime. The tangent bundle exponential map $EXP_p$ is smooth on $T_pM\times T_pM\setminus A$ with 
\begin{equation}
A=\{(U,V)\in T_pM\times T_pM | V\in (B\cup 0)\cap T_pM\}\,.
\end{equation}}

\vspace{6pt}\noindent\textit{Proof of Theorem 7.}
According to Theorem 3 the set $A$ is the set where the nonlinear connection coefficients $N^a{}_b$ which define the horizontal autoparallels of the Berwald covariant derivative are not smooth. For the Cartan nonlinear connection this is the set where the $L$ metric is degenerate. From the definition of Finsler spacetimes this is the zero section in $TM$ and the set $B$. $\square$

We like to point out here that for the application in physics the existence of the set $A\neq \emptyset$ is not problematic at all since there one is mostly interested in the geometry along timelike and lightlike directions $y$. There the geometry of spacetime, observers and light trajectories can be analyzed with help of the standard locally autoparallel coordinates, since there the $L$ metric is non-degenerate. 

Next we connect Finsler geodesics with special horizontal autoparallels of the Berwald linear connection to conclude that locally they can be identified with straight lines.

\vspace{6pt}\noindent\textbf{Theorem 8.}
\textit{Let $N^a{}_b$ be the connection coefficients of the Cartan nonlinear connection of a Finsler spacetime $(M,L)$. The horizontal autoparallels $ \gamma_{(U,U)}(t)$ of the Berwald connection are natural lifts of Finsler geodesics to the tangent bundle.}

\vspace{6pt}\noindent\textit{Proof of Theorem 8.}
By comparing the Finsler geodesic equation (\ref{eq:finsgeod}) and the equation which defines autoparallels of a general connection (\ref{eq:nlautopara0}) it is clear that Finsler geodesics are the autoparallels of the Cartan nonlinear connection. Theorem 2 of section \ref{sec:conn} states that the horizontal autoparallels $ \gamma_{(U,U)}(t)$ of the Berwald connection are autoparallels of the homogeneous and symmetric connection in consideration. Thus the statement of Theorem 8 holds. $\square$

\noindent From the definition of the exponential map in section \ref{sec:tmexp} and the first point of Theorem 4 Finsler geodesics can be one to one identified with special straight lines in $T_pM\times T_pM$. 

\vspace{6pt}\noindent\textbf{Corollary 1.}
\textit{The exponential map $EXP_p$ at $p=x_0\in M$ on a Finsler sapcetime  $(M,L)$ maps $(t U,U)\in T_pM\times T_pM\setminus A$ to the Finsler geodesic with initial data $x_U(0)=x_0$ and $\dot x_U(0)=U$ 
\begin{equation}
EXP_P(tU,U)=\gamma_{(tU,U)}(1)=\gamma_{(U,U)}(t)=(x_U(t),\dot x_U(t))\,.
\end{equation}}

\noindent As a remark for completeness we like to stress that in Finslerian geometry there exist several linear connections associated to the Cartan nonlinear connection, see \cite{FLG} or \cite{Minguzzi1} for a recent review of the topic. Instead of considering the horizontal autoparallels of the Berwald nonlinear connection one could equivalently use the horizontal autoparalles of the Cartan linear connection to define the exponential map. The horizontal autoparallels of the Cartan linear connection are defined similarly as the horizontal autoparallels of the Berwald connection, Definition 2, by the interchange of $\bar\partial_c N^a{}_b$ with 
\begin{equation}
\Gamma^{\delta a}{}_{bc}=\frac{1}{2}g^{L aq}(\delta_bg^L_{qc}+\delta_cg^L_{qb}-\delta_qg^L_{bc})\,,
\end{equation}
and obtaining a change in the autoparallel equation (\ref{eq:fauto}) to
\begin{equation}
\ddot x_{U}^a +  \Gamma^{\delta a}{}_{bc}(x_{U} , y_{V}))\dot x_{U}^b \dot x_{U}^c=0\,,
\end{equation}
However all proven properties of the exponential map and the locally autoparallel coordinates still hold in this case. The reason is that by the homogeneity of $N^a{}_b$ and the properties of the delta Christoffel symbols
\begin{equation}
\Gamma^{\delta a}{}_{bc}(x_{U} , y_{V}))y_V^b=N^a{}_b(x_U,y_V)=\bar\partial_bN^a{}_{c}(x_U,y_V)y_V^b\,.
\end{equation}
Thus in Finsler geometry there exist more than one possibility to define tangent bundle exponential maps which define locally autoparallel coordinates. However they differ only at quadratic order in $\tilde x$ in the expansion of the manifold induced coordinates $(x,y)$ with respect to the autoparallel coordinates $(\tilde x,\tilde y)$ in equations (\ref{eq:trafo1}) and (\ref{eq:trafo2}). In this article we stick to the tangent bundle exponential map we defined in Definition 3.

Next we calculate the component of geometric objects on Finsler spacetimes in locally autoparallel coordinates.

\subsection{Extended and standard locally autoparallel coordinates on Finsler spacetimes}
In section \ref{sec:lacs} we have seen that the connection coefficients in extended and standard locally autoparallel coordinates around $T_pM\subset TM$ vanish in $T_pM$. Both coordinate systems can be used in a Finslerian geometry since the Cartan nonlinear connection is homogeneous and symmetric. In virtue of Coroallary 1 it is justified to call the locally autoparallel coordinates in the context of a Finslerian geometry locally Finsler geodesic coordinates, since they use, among more general curves, Finsler geodesics as coordinate lines. Here we express the Finsler Lagrangian $L$ on a Finsler spacetime $(M,L)$ in both coordinate systems introduced in section \ref{sec:lacs} and the curvature of the Cartan nonlinear connection only in standard locally autoparallel coordinates. 

Recall that the extended locally autoparallel coordinates and the manifold induced coordinates are related by equations (\ref{eq:trafo1}) to (\ref{eq:trafo2}) and the coordinate transformation matrices are displayed in equations (\ref{eq:cm1}) to (\ref{eq:cm8}). The geometric objects we are interested in are all derived from derivatives of the Finsler Lagrangian $L$ and contain at most second derivatives with respect to $x$. We introduce the notation
\begin{equation}\label{eq:tildeL}
\tilde L(\tilde x, \tilde y)=L(x(\tilde x, \tilde y), y(\tilde x, \tilde y)),\ L(x,y)=\tilde L(\tilde x(x, y), \tilde y(x, y))\,,
\end{equation}
and express various derivatives acting on $\tilde L$ in terms of derivatives acting on $L$. Our aim is to expand $\tilde L$ in a power series with respect to $\tilde x$ to second order in both locally autoparallel coordinate systems
\begin{equation}\label{eq:Lexpand}
\tilde L(\tilde x, \tilde y)=\tilde L(0, \tilde y)+\tilde \partial_a\tilde L(0,\tilde y)\tilde x^a+\frac{1}{2}\tilde\partial_a\tilde\partial_b\tilde L(0,\tilde y)\tilde x^a\tilde x^b+\mathcal{O}(\tilde x^3)\,.
\end{equation}

\vspace{6pt}\noindent\textbf{Theorem 9.}
\textit{Let $\tilde L(\tilde x, \tilde y)$ be  the Finsler Lagrangian of a Finsler spacetime $(M, L)$ expressed in extended locally autoparallel coordinates. The $\tilde x$ derivatives of $\tilde L$ at $\tilde x=0$ are given by
\begin{equation}
\tilde L(0, \tilde y)=L(x_0,\tilde y),\ \tilde \partial_a \tilde L(0, \tilde y)=0,\ \tilde\partial_b \tilde\partial_a \tilde L(0, \tilde y)=0\,.
\end{equation}}

\vspace{6pt}\noindent\textit{Proof of Theorem 9.}
The first equation of the theorem can be read off from equation (\ref{eq:tildeL}). The second equality requires the fact that $L$ is horizontally constant, i.e. $\delta_aL=0$. With help of the coordinate transformation formulae (\ref{eq:cm1}) to (\ref{eq:cm4}) we equate
\begin{eqnarray}
\tilde \partial_a\tilde L(\tilde x,\tilde y)&=&\partial_m L(x(\tilde x, \tilde y), y(\tilde x, \tilde y))\tilde \partial_a x^m\nonumber\\
&+&\bar\partial_m L(x(\tilde x, \tilde y), y(\tilde x, \tilde y))\tilde \partial_a y^m\\
\Rightarrow\tilde \partial_a\tilde L(0,\tilde y)&=&\partial_a L(x_0,\tilde y )-N^m{}_a(x_0,\tilde y)\bar\partial_m L(x_0, \tilde y)\\
&=&\delta_aL(x_0,\tilde y)=0\,.
\end{eqnarray}
The third equality of the theorem is derived in a similar fashion but requires a bit more work
 \begin{eqnarray}
\tilde \partial_b\tilde \partial_a\tilde L(\tilde x,\tilde y)&=&\partial_n\partial_m L(x(\tilde x, \tilde y), y(\tilde x, \tilde y))\tilde \partial_a x^m\tilde \partial_b x^n\nonumber\\
&+&\bar\partial_n\partial_m L(x(\tilde x, \tilde y), y(\tilde x, \tilde y))\tilde \partial_a x^m\tilde \partial_b y^n\nonumber\\
&+&\partial_m L(x(\tilde x, \tilde y), y(\tilde x, \tilde y))\tilde\partial_b\tilde \partial_a x^m\nonumber\\
&+&\bar\partial_m L(x(\tilde x, \tilde y), y(\tilde x, \tilde y))\tilde\partial_b\tilde \partial_a y^m\nonumber\\
&+&\partial_n\bar\partial_m L(x(\tilde x, \tilde y), y(\tilde x, \tilde y))\tilde \partial_a y^m\tilde \partial_b x^n\nonumber\\
&+&\bar\partial_n\bar\partial_m L(x(\tilde x, \tilde y), y(\tilde x, \tilde y))\tilde \partial_a y^m\tilde \partial_b y^n\,,\\
\Rightarrow \tilde \partial_b\tilde \partial_a\tilde L(0,\tilde y)&=&\partial_a\partial_b L(x_0, \tilde y)-\bar\partial_n\partial_a L(x_0, \tilde y) N^{n}{}_b(x_0, \tilde y)\nonumber\\
&-&\partial_m L(x_0,\tilde y)\bar\partial_aN^{m}{}_b(x_0,\tilde y)-\bar\partial_m L(x_0, \tilde y)(\delta_{(a}N^m{}_{b)}(x_0,\tilde y)\nonumber\\
&-& N^m{}_r(x_0,\tilde y)\bar\partial_{a}N^r{}_{b}(x_0, \tilde y))\nonumber-\partial_b\bar\partial_m L(x_0, \tilde y) N^{m}{}_a(x_0, \tilde y)\nonumber\\
&+&\bar\partial_n\bar\partial_m L(x_0, \tilde y) N^{m}{}_a(x_0, \tilde y)N^{n}{}_b(x_0, \tilde y)\,.
\end{eqnarray} 
As in the previous calculation we use the fact that $L$ is horizontally constant
\begin{eqnarray}
0&=&\delta_a\delta_b L=\delta_a(\partial_b L-N^r{}_{b}\bar\partial_r L)\nonumber\\
&=&\partial_a\partial_b L-N^q{}_{a}\bar\partial_q\partial_b L-\delta_aN^r{}_b\bar\partial_r L-N^r{}_b\partial_a\bar\partial_r L+N^r{}_bN^q{}_a \bar\partial_q\bar\partial_r L\,,
\end{eqnarray}
and thus also $\partial_a L=N^q{}_a\bar\partial_q L$, to derive the equality
\begin{eqnarray}\label{eq:Rpulldowncontract}
\delta_aN^r{}_b\bar\partial_r L&=&\delta_a(N^r{}_b\bar\partial_r L)-N^r{}_b\delta_a\bar\partial_r L\nonumber \\
&=&\partial_a\partial_b L-N^r{}_a\bar\partial_r\partial_b L-N^r{}_b\partial_a\bar\partial_r L+N^r{}_aN^q{}_b\bar\partial_q\bar\partial_p L\nonumber\\&=&\delta_bN^r{}_a\bar\partial_r L \Rightarrow (\delta_aN^r{}_b-\delta_bN^r{}_a)\bar\partial_r L=R^r{}_{ab}\bar\partial_r L=0\,.
\end{eqnarray}
We obtain the desired result
\begin{eqnarray}
\tilde \partial_b\tilde \partial_a\tilde L(0,\tilde y)&=&\delta_aN^r{}_b(x_0,\tilde y)\bar\partial_r L(x_0,\tilde y)-\partial_m L(x_0,\tilde y)\bar\partial_aN^{m}{}_b(x_0,\tilde y)\nonumber\\
&-&\bar\partial_m L(x_0, \tilde y)[\delta_{(a}N^m{}_{b)}(x_0,\tilde y)- N^m{}_r(x_0,\tilde y)\bar\partial_{a}N^r{}_{b}(x_0, \tilde y)]\nonumber\\
&=&\frac{1}{2}\bar\partial_m L(\delta_aN^m{}_b-\delta_b N^m{}_a)-\delta_m L \bar\partial_aN^m{}_b=0\,.\ \square
\end{eqnarray}

\noindent Using the results of Theorem 9 in equation (\ref{eq:Lexpand}) yields that the Finsler Lagrangian does at most depend on $\tilde x$ in third order of its expansion.

\vspace{6pt}\noindent\textbf{Corollary 2.}
\textit{Let $\tilde L(\tilde x, \tilde y)$ be the Finsler Lagrangian of a Finsler spacetime $(M, L)$ in extended locally autoparallel coordinates. Its Taylor expansion around $\tilde x=0$ is
\begin{equation}
\tilde L(\tilde x, \tilde y)=L(x_0,\tilde y)+\mathcal{O}(\tilde x^3)\,.
\end{equation}}

\noindent For the standard locally autoparallel coordinates the result is different. They are related to manifold induced coordinates by the equations (\ref{eq:trafo1}) and (\ref{eq:trafo5}) which mimic the behaviour of a manifold induced coordinate change. Expanding the Finsler Lagrangian in $\tilde x$ we obtain an expression which is much more familiar to the expansion of the metric in metric geometry in normal coordinates.

\vspace{6pt}\noindent\textbf{Theorem 10.}
\textit{Let $\tilde L(\tilde x, \tilde y)$ be  the Finsler Lagrangian of a Finsler spacetime $(M, L)$ expressed in standard locally autoparallel coordinates. The $\tilde x$ derivatives of $\tilde L$ at $\tilde x=0$ are given by
\begin{equation}
\tilde L(0, \tilde y)=L(x_0,\tilde y),\ \tilde \partial_a \tilde L(0, \tilde y)=0,\ \tilde\partial_b \tilde\partial_a \tilde L(0, \tilde y)=\frac{2}{3}\tilde y^dR_{abd}(x_0,\tilde y)\,,
\end{equation}
where $R^{}_{abd}=g^L_{am}R^{m}{}_{bd}$ and $R^{m}{}_{bd}$ is the curvature of the Cartan nonlinear connection.}

\noindent Observe that the symmetry of the second derivative of $\tilde L$ with respect to $\tilde x$ is given by the identity $R_{[abd]}=0$, see for example \cite{FLG}.

\vspace{6pt}\noindent\textit{Proof of Theorem 10.}
The first two equalities of Theorem 10 are proven in exactly the same way as the first two equalities of Theorem 9, since they only require the first order expansion of the manifold induced coordinates in standard locally autoparallel coordinates. These are identical to the expansion in extended locally autoparallel coordinates. To proof the third equality we need the second order expansion of $y(\tilde x, \tilde y)$ in $\tilde x$ in the standard locally autoparallel coordinates which differs from the expansion in extended locally autoparallel coordinates. First of all we use again the horizontality of $L$ to write
\begin{eqnarray}
\tilde \partial_a\tilde L(\tilde x,\tilde y)&=&\partial_m L(x(\tilde x, \tilde y), y(\tilde x, \tilde y))\tilde \partial_a x^m+\bar\partial_m L(x(\tilde x, \tilde y), y(\tilde x, \tilde y))\tilde \partial_a y^m\nonumber\\
&=&\bar\partial_m L(x(\tilde x, \tilde y), y(\tilde x, \tilde y))(\tilde\partial_a y^m+N^m{}_r(x(\tilde x, \tilde y), y(\tilde x, \tilde y)\tilde\partial_ax^r)\,.
\end{eqnarray}
We already know that $Q^m{}_a(\tilde x, \tilde y)=\tilde\partial_a y^m+N^m{}_r(x(\tilde x, \tilde y), y(\tilde x, \tilde y)\tilde\partial_ax^r$ vanishes at $\tilde x=0$, since $\tilde \partial_a\tilde L(0,\tilde y)=0$, so that we only need to consider the terms including derivatives acting on $Q^m{}_a$ for further derivatives acting on $\tilde L$. Thus
\begin{eqnarray}\label{eq:ddtildel}
\tilde \partial_b\tilde \partial_a\tilde L(0,\tilde y)=\bar\partial_m L(x_0, \tilde y)\tilde\partial_bQ^m{}_a(0,\tilde y)
\end{eqnarray}
From the definition of $Q^m{}_a(0,\tilde y)$ and the coordinate transformation formulae (\ref{eq:cm1}) to (\ref{eq:cm4}), so far we only needed them to first order, it is straightforward to equate
\begin{equation}
\tilde\partial_bQ^m{}_a(0,\tilde y)=\tilde\partial_b\tilde\partial_a y^m(0,\tilde y)+N^m{}_r(x_0, \tilde y)\tilde\partial_b\tilde\partial_ax^r(0,\tilde y)+\delta_a N^m{}_b(x_0,\tilde y)\,.
\end{equation}
The second derivatives of the coordinates $x$ and $y$ with respect to $\tilde x$ can be read off from equations (\ref{eq:cm1}) and (\ref{eq:trafo5})
 \begin{eqnarray}
\tilde\partial_bQ^m{}_a(0,\tilde y)&=&-(\delta_{(d}\bar\partial_aN^m{}_{b)}(x_0,\tilde y)-2\bar\partial_rN^m{}_{(d}(x_0,\tilde y)\bar\partial_{b}N^r{}_{a)}(x_0, \tilde y))\tilde y^d\nonumber\\
&-&N^m{}_r(x_0, \tilde y)\bar\partial_aN^r{}_{b}(0,\tilde y)+\delta_a N^m{}_b(x_0,\tilde y)\,.
\end{eqnarray}
Expanding the symmetrisation brackets and inserting by homogeneity 
\begin{eqnarray}
\delta_a N^m{}_b(x_0,\tilde y)&=&[\delta_a (y^d\bar\partial_dN^m{}_b)](x_0,\tilde y)\nonumber\\
&=&\tilde y^d\delta_a\bar\partial_dN^m{}_b(x_0,\tilde y)-N^d{}_a(x_0,\tilde y)\partial_dN^m{}_b(x_0,\tilde y)
\end{eqnarray}
as well as $N^m{}_r(x_0, \tilde y)\bar\partial_aN^r{}_{b}(x_0,\tilde y)=\tilde y^d\bar\partial_dN^m{}_r(x_0, \tilde y)\bar\partial_aN^r{}_{b}(x_0,\tilde y)$ yields 
\begin{eqnarray}
&&\tilde\partial_bQ^m{}_a(0,\tilde y)\nonumber\\
&=&\frac{1}{3}\tilde y^d[\delta_b\bar\partial_dN^m{}_a(x_0,\tilde y)-\delta_d\bar\partial_bN^m{}_a(x_0,\tilde y)\nonumber\\
&+&\bar\partial_rN^m{}_b(x_0,\tilde y)\bar\partial_aN^r{}_d(x_0,\tilde y)-\bar\partial_rN^m{}_d(x_0,\tilde y)\bar\partial_aN^r{}_b(x_0,\tilde y)]\nonumber\\
&+&\frac{1}{3}\tilde y^d[\delta_b\bar\partial_dN^m{}_a(x_0,\tilde y)-\delta_a\bar\partial_bN^m{}_d(x_0,\tilde y)\nonumber\\
&+&\bar\partial_rN^m{}_b(x_0,\tilde y)\bar\partial_aN^r{}_d(x_0,\tilde y)-\bar\partial_rN^m{}_a(x_0,\tilde y)\bar\partial_bN^r{}_d(x_0,\tilde y)]\nonumber\\
&=&\frac{1}{3}\tilde y^d(\bar\partial_aR^m{}_{db}(x_0,\tilde y)+\bar\partial_dR^m{}_{ab}(x_0,\tilde y))\nonumber\\
&=&\frac{1}{3}(\tilde y^d\bar\partial_aR^m{}_{db}(x_0,\tilde y)+R^m{}_{ab}(x_0,\tilde y))\,.
\end{eqnarray}
Together with equation (\ref{eq:ddtildel}) and $ R^m{}_{ab}\bar\partial_mL=0$ (see equation (\ref{eq:Rpulldowncontract}) from the proof of the previous theorem) and the introduction of $R_{abd}=g^L_{am}R^m{}_{bd}$ this proofs the last equality of Theorem 10. $\square$

\noindent Theorem 10 leads directly to the expansion of $\tilde L(\tilde x, \tilde y)$ in a power series with respect to $\tilde x$ in the standard locally autoparallel coordinates.

\vspace{6pt}\noindent\textbf{Corollary 3.}
\textit{Let $\tilde L(\tilde x, \tilde y)$ be Finsler Lagrangian of a Finsler spacetime $(M, L)$ in standard locally autoparallel coordinates. Its taylor expansion around $\tilde x=0$ is
\begin{equation}
\tilde L(\tilde x, \tilde y)=L(x_0,\tilde y)+\frac{1}{3} \tilde y^dg^{L}_{am}(x_0,\tilde y)R^m{}_{bd}(x_0,\tilde y)\tilde x^a\tilde x^b+\mathcal{O}(\tilde x^3)\,.
\end{equation}}
Observe that the expansion of the Finsler Lagrangian Corollary above becomes the usual expansion of the metric in Riemann normal coordinates when a metric Finsler Lagrangian $L=g_{ab}(x)y^ay^b$ and a change of basis from manifold induced coordinate basis to a frame of the metric $g=g_{ab}(x)dx^a\otimes dx^b$ is considered.

As last feature in this article we like to express the components of the nonlinear curvature tensor of the Cartan nonlinear connection in the standard autoparallel coordinates. Observe that the second derivative of $L$ with respect to $\tilde x$ obeys the following identity, which is derived from Corollary 3 above
\begin{eqnarray}
\tilde y^m\tilde\partial_m\tilde\partial_b\tilde{\bar\partial}_a\tilde L(0,\tilde y)=-\tilde\partial_a\tilde\partial_b\tilde L(0,\tilde y)=-\frac{2}{3}\tilde y^dR_{abd}(x_0,\tilde y)\,.
\end{eqnarray}
Thus the first $\tilde x$ derivative of the the nonlinear connection coefficients in the reduced locally autoparallel coordinates is given by
\begin{equation}
\tilde\partial_c\tilde N^a{}_b(0,\tilde y)=\frac{1}{4}\tilde{\bar\partial}_b[\tilde g^{qa}(y^m\tilde \partial_m\tilde\partial_c\tilde{\bar\partial}_q\tilde L-\tilde\partial_c\tilde\partial_q \tilde L)]=-\frac{1}{2}\tilde{\bar\partial}_b[\tilde g^{qa}\tilde\partial_c\tilde\partial_q \tilde L]
\end{equation}
and the components of the curvature of the Cartan nonlinear connection become just a few derivatives action on the Finsler Lagrangian~$\tilde L$
\begin{eqnarray}
\tilde R^a{}_{bc}(0,\tilde y)&=&\tilde \delta_c\tilde N^a{}_{b}-\tilde \delta_b\tilde N^a{}_{c}=\tilde \partial_c\tilde N^a{}_{b}-\tilde \partial_b\tilde N^a{}_{c}\nonumber\\
&=&\frac{1}{2}\tilde{\bar\partial}_c\tilde [g^{qa}\tilde\partial_b\tilde\partial_q \tilde L]-\frac{1}{2}\tilde{\bar\partial}_b[\tilde g^{qa}\tilde\partial_c\tilde\partial_q \tilde L]\,.
\end{eqnarray}
Moreover from here it is straightforward to express the the components $\tilde R^a{}_{bc}$ in terms of their counterpart in manifold induced coordinates. With $\bar\partial_{[b}R^a{}_{cd]}=0$ we equate
\begin{eqnarray}
&&\tilde R^a{}_{bc}(0,\tilde y)\nonumber\\
&=&\frac{1}{3}(\tilde{\bar\partial}_c(\tilde y^dR^a{}_{bq})-\tilde{\bar\partial}_b(\tilde y^dR^a{}_{cq}))=\frac{1}{3}(R^a{}_{bc}-R^a{}_{cb}+\tilde y^d (\tilde {\bar\partial}_c R^a{}_{bd}-\tilde{\bar\partial}_bR^a{}_{cd}))\nonumber\\
&=&R^a{}_{bc}(x_0,\tilde y)\,.
\end{eqnarray}
The last equality demonstrates also that the components of the curvature of the Cartan nonlinear connection in standard locally autoparallel coordinates do not change compared to manifold induced coordinates even though a not manifold induced coordinate transformation was applied.

\section{Discussion}\label{sec:disc}
Throughout this article we constructed two autoparallel coordinate systems for general connections on the tangent bundle. The key insight for the construction of the coordinates is that they are coordinates on the tangent bundle of the manifold in consideration and not on the manifold itself. Their key features are the vanishing of the connection coefficients of the general connection at the origin of the new coordinate system, as stated in Theorem~6, and that the horizontal autoparallels of the  Berwald linear and even the autoparallels of a homogeneous and symmetric connection become straight lines, as stated in Theorem~4. For Finslerian geometries one of the coordinate systems, the standard locally autoparallel coordiantes become the Douglas-Thomas autoparallel coordinates, and Finslerian geodesics become straight lines, as concluded in Theorem 8. Moreover we expanded the Finsler Lagrangian $L$ of a Finsler spacetimes in a power series having earliest quadratic dependence on the new coordinates along the manifold in Corollary~2 and Corollary~3. The strong result from Corollary~2 demonstrates very clearly that the extended locally autoparallel coordinates use the whole freedom of the tangent bundle, i.e. they introduce new position coordinates $\tilde x$ and direction coordinates $\tilde y$ independently from each other.  This freedom leads to a dependence of the Finsler Lagrangian on the new coordinates which is simpler than one ever could achieve using only coordinate transformations on the manifold. It is an open question how these coordinates can be interpreted physically, since even for metric spacetime geometry they enable us to transform the curvature away. For the standard locally autoparallel coordinates the new direction coordinates $\tilde y$ are derived from the change of the position coordinates $\tilde x$ in a way such that they mimic the behaviour of a manifold induced coordinate change on the tangent bundle. As in metric geometry the locally autoparallel coordinates lead to massive simplifications in the expression of geometric tensors in Finslerian geometry such as the components of the curvature. The construction of the coordinate systems adapted to the geometry of a manifold beyond affine connection geometry during this article shows that the geometry of a manifold defined by a general connection on the manifold is not so different to the well known affine connection geometry. In particular the application to Finslerian geometries demonstrates that the general Finsler geometric framework is less different to the metric geometry framework as sometimes expressed.

For physics the existence of standard locally autoparallel coordinates may lead to an interesting interpretation. In general relativity the existence of normal coordinates in metric geometry is interpreted as the description of an Einstein elevator, i.e. as a small freely falling laboratory in which free particles move along straight lines as in special relativity. For the application of Finsler geometry as extended geometry of spacetime in the future the standard locally autoparallel coordinates can be investigated towards the interpretation as an Einstein elevator on Finsler spacetimes.

 \section*{Acknowledgements} I thank Volker Perlick and Manuel Hohmann for inspiring discussions and remarks and I gratefully acknowledge the financial support as Riemann Fellow from the Riemann Center for Geometry and Physics of the Leibniz University Hannover.

\appendix
\section{Transformation of a connection}\label{app:nlintrafo}
Let $\omega$ be a connection on $TM$. In manifold induced coordinates $(x, y)$ it has the form
\begin{equation}
\omega=\omega_{(x,y)}=(dy^a+N^a{}_b(x,y)dx^b)\otimes \bar\partial_a\,.
\end{equation}
A coordinate transformation to arbitrary new coordinates $(\tilde x, \tilde y)$ changes $\omega$ to
\begin{eqnarray}
\omega_{(x(\tilde x,\tilde y),y(\tilde x,\tilde y))}\nonumber
&=&\bigg[\frac{\partial y^a}{\partial \tilde y^b}\frac{\partial \tilde x^q}{\partial y^a}+N^a{}_i\frac{\partial x^i}{\partial \tilde y^b}\frac{\partial \tilde x^q}{\partial y^a}\bigg]d\tilde y^b\otimes\tilde\partial_q\\
&+&\bigg[\frac{\partial y^a}{\partial \tilde y^b}\frac{\partial \tilde y^q}{\partial y^a}+N^a{}_i\frac{\partial x^i}{\partial \tilde y^b}\frac{\partial \tilde y^q}{\partial y^a}\bigg]d\tilde y^b\otimes\tilde{\bar\partial}_q\nonumber\\
&+&\bigg[\frac{\partial y^a}{\partial \tilde x^b}\frac{\partial \tilde x^q}{\partial y^a}+N^a{}_i\frac{\partial x^i}{\partial \tilde x^b}\frac{\partial \tilde x^q}{\partial y^a}\bigg]d\tilde x^b\otimes\tilde\partial_q\nonumber\\
&+&\bigg[\frac{\partial y^a}{\partial \tilde x^b}\frac{\partial \tilde y^q}{\partial y^a}+N^a{}_i\frac{\partial x^i}{\partial \tilde x^b}\frac{\partial \tilde y^q}{\partial y^a}\bigg]d\tilde x^b\otimes\tilde{\bar\partial}_q\,.
\end{eqnarray}
For a manifold induced coordinate transformation $\tilde x =\tilde x (x)$ and thus $\tilde{\bar\partial}_ax^b=0$ and $\bar\partial_a\tilde x^b=0$ one obtains the transformation equation (\ref{eq:ntrafo}). To obtain formula (\ref{eq:tildenvanish}) used to proof Theorem 6, simply insert the coordinate expansion in equation (\ref{eq:cm1}) to (\ref{eq:cm4}) and evaluate at the point $(\tilde x, \tilde y)=(0, \tilde y)$.

\bibliographystyle{utphys}
\bibliography{testFNC}

\end{document}